\documentclass[12pt]{article}
\usepackage{mathtext}
\usepackage{graphicx}
\usepackage{amsfonts}
\usepackage{amssymb}
\usepackage{amsmath}
\usepackage{pgfplots}
\usepackage{braket}
\usepackage{tikz}
\usepackage{float}
\newcommand{\la}{\lambda}
\renewcommand{\a}{\alpha}
\begin{document}
\title{Quantum gates on asynchronous atomic excitations}

\author{Yuri Ozhigov$^{1,2}$,\thanks{e-mail: ozhigov@cs.msu.su} \\
{\it 
1. Lomonosov Moscow State University, 
} \\
{\it Faculty of Computational Mathematics and Cybernetics,}
\\
{\it 2. Valiev Institute of Physics and Technology }
\\
{\it Russian Academy of Sciences, Moscow, Russia}
\\ 
}
\maketitle

\begin{abstract}
The article proposes the implementation of a universal system of quantum gates on asynchronous excitations of two-level atoms in optical cavities. The entangling operator of the CSign type is implemented without beam splitters, approximately, based on the incommensurability of the periods of Rabi oscillations in a cavity with single and double excitations.
\end{abstract}
\section{Introduction}

Quantum computing is an actual intrusion of quantum theory into the field of complex processes, where the action of its basic laws has not yet been studied. Therefore, the construction of the simplest schemes of such computations, in which quantum laws would appear as clearly as possible, is still an urgent task. The dark place here is decoherence, which occurs due to the interaction of charges and the field, the quanta of which closely link the quantum computer with the environment. This makes it necessary to account for and control, or even explicitly use photons in quantum computing.

Photons as data carriers make it possible to use linear optical devices to implement one qubit gates, but the construction of entangling operations is difficult, since photons do not directly interact with each other. There is a popular KLM scheme (see \cite{KLM}) where measurements are used as an interaction ersatz, and its improvement with teleportation (see \cite{DC}), which significantly increases its efficiency, and some variants of this scheme for atoms (see, for example, \cite{Po}). However, the use of classical probabilistic schemes in practical implementation places increased demands on the efficiency of quantum gates on single particles. Using classical probability obscures the main question of a quantum computer: how does coherence work for complex systems with different particles?

The most basic methods are more suitable here, the main one being optical cavities with several atoms, whose interaction with a single-mode field is clearly described from the first principles (for the capabilities of this type of device, see, for example, \cite{Re}). Thus, the CNOT gate was implemented using the external - vibrational - degrees of freedom of the atom (see \cite{Wi}). However, the essence of quantum computing is not in the coherent behavior of a single qubit, but in the scaling of a Feynman quantum processor that implements the theoretical possibilities of unitary dynamics in the entire Hilbert space of states, giving, for example, Grover algorithm (\cite{G}) on the same hardware as Shor algorithm (\cite{Sh}). Using external factors to demonstrate the dynamics of individual atoms and the field is useful for individual atoms, but the inevitable disturbance caused by these factors will affect scalability.

Therefore, the gate implementation schemes that use minimal tools and can be described from the first principles have value. One of these schemes was proposed by H. Azuma in \cite{A}, where dual-rail states of single photons are used as qubits. In this scheme the interaction of photons with atoms is only used to perform the entangling $CSign$ transformation, which requires two optical cavities, two beam splitters and phase shifters.

This article proposes a simplification of the Azuma scheme, where only one cavity is used, and the beam splitters are replaced by a time shift for the photons entering it. The logical qubits are  asynchronous states of the atoms in Rabi oscillations. This scheme can be modified for purely photonic carriers, with a time shift that determines the qubit value. However, atoms as information carriers have the advantage that they are much easier to control, as well as the photons they emit. The advantage of the proposed scheme is its simplicity. The disadvantage is the same as in the  scheme from \cite{A} - the dependence on the Pockels cell (or its analogue) reaction time that must be significantly less than the time of the Rabi oscillations of the atom in the cavity.

For technical reasons, we will implement the gate $coCSign:\ |x,y\rangle\rightarrow (-1)^{(x\oplus 1)y} |x,y\rangle$, which changes the sign at the single state $|01\rangle$, which is related to the gate $CSign:\ |x, y \rangle\rightarrow (-1)^{xy}|x, y \rangle$, implemented in \cite{A}; in fact it's the same thing, since$CSign=\sigma_x(x)\ coCSign\ \sigma_x(x)$ and one qubit gates are implemented by linear optical devices.

\section{Calculation of phase shifts}

The core of this scheme is an optical cavity with a single two-level atom with an energy gap $ \hbar\omega$ between the main $|0 \rangle$ and the excited $|1\rangle$ levels, where $\omega$ matches the frequency of a photon of a certain mode held in the cavity. The interaction constant $g$ between the atom and the field is assumed to be small: $g/\hbar\omega\ll 1$ (in practice, this ratio must be no more than $10^{-3}$) for the possibility of applying the rotating wave approximation (RWA), in which the Jaynes-Cummings Hamiltonian of the system ''atom+field''  (\cite{JC}) has the form
\begin{equation}
H=H_{JC}=H_0+H_{int};\ H_0=\hbar\omega a^+a+\hbar\sigma^+\sigma,\ H_{int}=g(\a^+\sigma+a\sigma^+),
\label{HamJC}
\end{equation}
where $a, a^+$ are the photon annihilation and creation operators, $\sigma,\sigma^+$ are the relaxation and excitation operators of the atom. We will write the basic states of the atom and the field as $|n\rangle_{ph}|m\rangle_{at}$, where $n=0,1,2,...$ is the number of photons in the cavity, $m=0,1$ is the number of atomic excitations. We will have $n=0,1,2$. The Hamiltonian during the execution of the $coCSign$ gate will change: a summand $H_{jump}=\nu (a_ia^+_j+a_ja^+_i)$ meaning the transition of a photon from the cavity $i $ to the cavity $j$ and vice versa will be added to $H_{int}$ that does not change the energy of independent atoms and field $H_0$. We thus obtain Jaynes-Cummings-Hubbard model (JCH). Therefore, the adding to the phase associated with $H_0$ will be the same for all states and can be ignored. Next, we will consider the adding to the phase relative to either the identical operator $I$, or to $\sigma_x$, since all the operations discussed below are reduced to either the first or the second case, so the adding to the phase when applied, for example, to $ - i\sigma_x$ will be $ - \frac {\pi}{2}$.

Let $\tau_1= \pi\hbar/g,\ \tau_2=\pi\hbar/g\sqrt{2}$ be the periods of Rabi oscillations:
$$
|n\rangle_{ph}|0\rangle_{at}\leftrightarrow |n-1\rangle_{ph}|1\rangle_{at}
$$
 in the cavity for the total energy $n\hbar\omega$, for $n=1$ or $n=2$ respectively. The evolution operator $U_t=e^{-\frac{i}{\hbar}Ht}$ at important times will depend on the total energy of the cavity. If the energy is equal to $ \hbar\omega$, in the basis $| \phi_0\rangle=|1\rangle_{ph}|0\rangle_{at}, \ | \phi_1\rangle=|0\rangle_{ph}|1\rangle_{at}$, we have:
\begin{equation}
U_{\tau_1/2}=-i\sigma_x,\ U_{\tau_1}=-I,\ U_{2\tau_1}=I,
\label{phase_add}
\end{equation}
where $\sigma_x$ is the Pauli matrix, and similar relations with the replacement of $\tau_1$ by $\tau_2$ for the total energy of the cavity $2\hbar\omega$.

Moving a photon from the cavity $j$ to the cavity $i$ and vice versa is realized by simultaneous turning on Pockels cells or similar devices in these cavities that is formally represented by the addition of $H_{jump}$ to the interaction $H_{int}$. In the absence of atoms in the cavities, this action implements exactly the same dynamics as Rabi oscillations, but with the period $\tau_{jump}=\pi\hbar/\nu$ provided that there is only one photon per two cavities. We will assume that $\nu\gg g$ and it is possible to move a photon from a cavity to a cavity so that the atom does not affect this process at all; the corresponding addition to the phase can be thus found by formulas similar to \eqref{phase_add}. As noted in \cite{A}, this is difficult to implement in an experiment, but there are reasons to consider it as a technical difficulty. If this condition is satisfied, the addition to the phase when moving a photon from one cavity to another will be $-\pi/2$, the same as for half of the Rabi oscillation.

Due to the incommensurability of the periods of Rabi oscillations $\tau_1$ and $\tau_2$, we can choose such natural numbers $n_1$ and $n_2$ that the following approximate equality will be fulfilled with high accuracy

\begin{equation}
2n_2\tau_2\approx 2n_1\tau_1+\frac{\tau_1}{2};
\label{noncommon}
\end{equation}
it will be the basis for the nonlinear phase shift required for implementation of $coCSing$. 

\section{Implementation of $coCSign$}

The state of the qubit $|0\rangle$ is implemented in our model as the state of the optical cavity of the form $|0\rangle_{ph}|1\rangle_{at}$, and the state of the qubit $|1\rangle$ - as $|1\rangle_{ph}|0\rangle_{at}$. Thus, for the state $|01 \rangle$, which needs to invert the phase, our representation has the form $|01\rangle_{ph}|10\rangle_{at}$, where the first photon qubit belongs to the cavity $x$ , and the second-to the cavity $y$. Note that after a time of $\tau_1/2$, the zero and the unit change places with the addition to phase $-\pi/2$.

\begin{figure}
\begin{centering}
\includegraphics[scale=0.60]{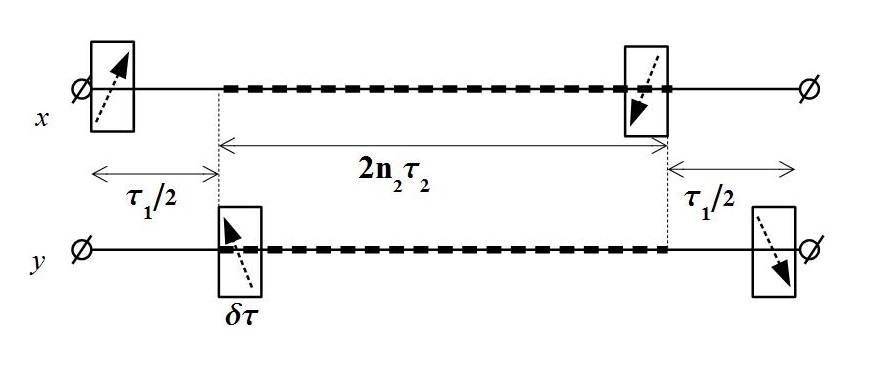}
\caption{Sequence of operations when implementing the gate coCSign: $|x,y\rangle\rightarrow (-1)^{x(y\oplus 1)}|x,y\rangle$ on asynchronous atomic excitations in optical cavities, $\delta\tau=\tau_{jump}/2$. After this scheme we have to wait for the time $\tau_1/2$.}
\label{fig:1}
\end{centering}
\end{figure}

\begin{figure}
\begin{centering}
\includegraphics[scale=0.60]{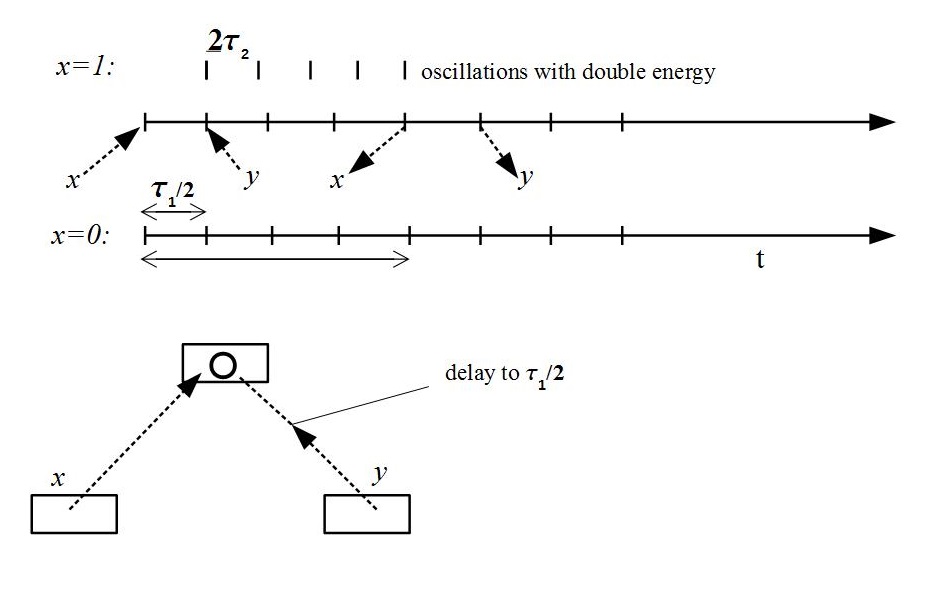}
\caption{Implementing the gate coCSign}
\label{fig:coCSign}
\end{centering}
\end{figure}

The sequence of operations that implement $coCSign$ is shown in figure \ref{fig:1}, and the participating cavities are shown in figure \ref{fig:coCSign}. First, we organize the exchange of photon between the auxiliary cavity with zero initial total energy and cavity $x$, then, with the delay $\tau_1/2$ - the similar photon exchange with cavity $y$, then after time $2n_2\tau_2$, again the photon echange between the auxiliary cavity and cavity $x$, then, after the time $\tau_1/2$ the similar photon exchange with cavity $y$. From our choice of photon travel times (see \eqref{noncommon}), it follows that at these moments there will be either one photon or none in the participating cavities, so all exchange operations on the short time frame $\delta \tau=\pi\hbar/2\nu\ll \tau_1$ will give exactly the movement of photons.

From the figure \ref{fig:1} it can be seen that the evolution of the state has three long segments (interaction of atoms and photons in cavities), and four short segments (jumps of a photon from a cavity to a cavity).
Consider in detail the phase change for the state $|00\rangle$.

1 segment, short. Nothing happens, since there is no photon in the "x" cavity and in the auxiliary one.

2 section, long. Joint evolution of 2 cavities, in each atom and photon; the addition to the phase is $ - \pi/2 -\pi/2$.

3 section, short. The photon in the cavity $y$, is transferred to the auxiliary cavity, the phase addition is $-\pi/2$.

4 section, long. Evolution of two cavities, in each energy $\hbar\omega$, phase addition is
$-\pi/2 -\pi/2$.

5 section, short. Nothing happens because there is no photon in the participating cavities.

6 section, long. Joint evolution of 2 cavities, in each atom and photon phase addition is $ -\pi/2 -\pi/2$.

7 section, short. The photon from the auxiliary cavity is transferred to the "y" cavity; the phase addition is $ -\pi/2$.

As a result, the total phase run is $ 8\pi/2 =0$.

In this case, the state $ |00\rangle$ goes to $ |11\rangle$, after waiting for $\tau_1/2$,  that is returns back with the addition to the phase $\pi$ - but this addition is common to all initial basic states, and therefore it can be ignored.

Next, the same scheme is used for transitions of the following type: $
|01\rangle \rightarrow |10\rangle$ - phase addition is $-\pi$; here it is provided by the equality \eqref{noncommon}, $ |11\rangle \rightarrow |00\rangle$, phase addition $0$, and finally $|10\rangle\rightarrow |01\rangle$, with phase addition $0$. After waiting for the time $ \tau_1 / 2$, all initial basic states take the same form, with a common phase addition $\pi$.

The article \cite{SOL} provides a calculation that implies that to achieve satisfactory accuracy of such entangling gates as $CSign$ on nonlinearity in cavities, it is sufficient to take the numbers of  incommensurable periods $n_1,n_2$ that do not exceed several tens, which corresponds to the number of observed Rabi oscillations in optical cavities.

So, the main difficulty is in the speed of the Pockels cells operation that is technically avoidable thing.

We have shown the possibility of replacing beam splitters with time shifts, which simplifies the implementation of entangling gates. 
The main advantage of the proposed gate implementation scheme is its simplicity and the ability to accurately follow the JCH theoretical model, which, despite the mentioned technical difficulty, inspires optimism about the scalability of the quantum computer and about the comparison of its theory with experiments on a large number of qubits.

\section{Physical restrictions on the quality of the gate coCSign}

The advantage of the above coCSign gate scheme is that, unlike, for example, the KLM scheme (\cite{KLM}), it is completely implemented by means of the standard Janes-Cummings-Hubbard model and, ideally, does not require any operations that go beyond it.
The article \cite{SOL} provides a calculation that shows that to achieve a satisfactory accuracy of such entangling gates on nonlinearity in cavities, it is sufficient to take the number of incommensurable periods $n_1, n_2$, not exceeding several tens, which corresponds to the number of observed Rabi oscillations in optical cavities.

However, the JCH scheme itself has restrictions on the parameters contained in it, so any arbitrary choice of their values may lead to a violation of its applicability.
In the article \cite{A}, one of these restrictions, which follows from the experiments, is noted - on the rate of activation of the Pokkels cell. But this limitation is not the only one.

The force of interaction of an atom with a field in the cavity has the form
\begin{equation}
g=\sqrt{\hbar\omega/V}d\ E(x),
\label{g}
\end{equation}
where $V$ is the effective volume of the cavity, $d$ is the dipole moment of transition between the main and perturbed states, $E(x)$ is the factor of the spatial location of the atom in the cavity, equal to $E(x)=sin(\pi x/L)$, where $L$ is the length of the cavity. For reliable retention of the photon in the cavity, we must have $L=n\la/2$, where $\la$ is the photon's wavelength; in experiments, we take $n=1$ to reduce the effective volume of the cavity, which allows us to obtain several tens of Rabi oscillations (see, for example,\cite{Re}). 

We cannot choose the parameter $\nu$ of the photon transition intensity to be too large due to the energy - time uncertainty relation for photons, since the too short period of time for a photon to pass from a cavity to a cavity automatically means a large uncertainty of its energy, and therefore leads to a rapid loss of the photon, whose wavelength becomes too different from twice the length of the cavity.

Given that the frequency of the photon for experiments with the rubidium atom is approximately $10^{10}\ c^{-1}$ and taking the upper bound of the possible uncertainty of the frequency $10^9\ c^{-1}$ (the real bound is much smaller), given the uncertainty ratio $ \delta\omega\ \delta t\approx 1$, we get a lower bound for the time window of the photon transition $\delta\tau\approx 10^{-9}\ s^{-1}$. Taking into account the time of the Rabi oscillation $\tau\approx 10^{-6}\ s^{-1}$, we obtain the inequality for the time window
$10^{-9}\leq\delta\tau\ll 10^{-6}$, which means that the gate can be triggered once with an error that exceeds $10^{-3}$. This error, unfortunately, does not make it possible to build a Feynman quantum computer on this processor, if we consider its only option.

We note that the above inequalities are only material for discussion, and in no case can serve as an estimate of the real error of this gate, since the uncertainty ratio acts together with other decoherence factors: the inaccuracy of the equality \eqref{noncommon} and the limited lifetime of the photon in the cavity.
For example, we could increase the range of possible changes in the magnitude of $\delta\tau$, reducing the period of Rabi oscillations, for example, by reducing the spatial location $E(x)$, but this way will reduce the lifetime of photon in cavity, which no less fatal to the gate.

This difficulty is found in all schemes of photon gates. The solution lies in the use of many processors, that is, in the combination of the quantum effect with the effect of classical parallelization; a similar technique is used in other well-known schemes of photonic computers, for example, in the already mentioned KLM scheme.

In the literature, there are no more accurate estimates of the combined action of three factors: inaccuracies in determining the time $n_2\tau_2$, errors due to the restriction on the time window of photon transfer, and the limitation of the number of oscillations. The proposed scheme is the simplest known, and all its disadvantages are typical for other similar schemes. It is possible to evaluate the suitability of photonic quantum systems for calculations only in an experiment, and the scheme considered is one of the best candidates for implementation.


\section{Acknowledgements}

The work is supported by the Russian Foundation for Basic Research, grant a-18-01-00695.

\newpage

\end{document}